# Estimation of Defect proneness Using Design complexity Measurements in Object-Oriented Software

R. Selvarani
*Computer Science & Engg.*
*RIIC, DS Institutions*
Bangalore, India
selvss@yahoo.com

T.R.Gopalakrishnan Nair
*Computer Science & Engg*
*RIIC, DS Institutions*
Bangalore, India
trgnair@yahoo.com

V. Kamakshi Prasad
*Computer Science & Engg*
*J.N.Technological University*
Hyderabad, India
kamakshiprasad@yahoo.com

***Abstract*—Software engineering is continuously facing the challenges of growing complexity of software packages and increased level of data on defects and drawbacks from software production process. This makes a clarion call for inventions and methods which can enable a more reusable, reliable, easily maintainable and high quality software systems with deeper control on software generation process. Quality and productivity are indeed the two most important parameters for controlling any industrial process. Implementation of a successful control system requires some means of measurement. Software metrics play an important role in the management aspects of the software development process such as better planning, assessment of improvements, resource allocation and reduction of unpredictability. The process involving early detection of potential problems, productivity evaluation and evaluating external quality factors such as reusability, maintainability, defect proneness and complexity are of utmost importance. Here we discuss the application of CK metrics and estimation model to predict the external quality parameters for optimizing the design process and production process for desired levels of quality. Estimation of defect-proneness in object-oriented system at design level is developed using a novel methodology where models of relationship between CK metrics and defect-proneness index is achieved. A multifunctional estimation approach captures the correlation between CK metrics and defect proneness level of software modules.**

## I. INTRODUCTION

In the object-oriented environment, one of the major aspects having strong influence on the quality of resulting software system is the design complexity. The structural property of the software component is influenced by the cognitive properties of the individuals involved in designing, development and testing, and it will be reflected in the structural properties of the developed software. The OO paradigm offers the technology to create components that can be used for generic programming. CK metrics suite [2] is one of the object-oriented design complexity measurement systems which support the measurement of the external quality parameter which may evolve in software package. The literature widely refers to the metric suite which depends on the internal structural analysis of object-oriented components such as inheritance, coupling, cohesion, method invocation, and association [ 4, 5, 6].

It is found that there exists a complex functional relationship between the internal design qualities and the defect proneness of the system. The model arrived in this work predicts the defect proneness index of software based on the three factors such as Depth of Inheritance (DIT), Response For a Class (RFC) and Weighted Method per Class (WMC) of CK metric suite. The method of implementation of this approach involves the analysis and estimation of functional relationship presenting the system properties which offers external parameters like defect proneness. A set of wider observation on real world values of external quality parameters with selected CK metrics have offered the relationship. Once relationship has been established through various functional estimation techniques the same relationship is used to predict defect-proneness index of classes obtained by evaluating CK Metrics. Table I depicts the definitions and NASA–Rosenberg thresholds of CK metrics [3].

TABLE I. CK METRICS

| Metric | NASA–Rosenberg Threshold [3] | Description |
|---|---|---|
| WMC | 20, Maximum 100 | Sum of complexities of methods in a class or Count of methods of a class. High value indicates decreased Reusability and Maintainability. |
| RFC | 50, 100 (Maximum 222) | It measures the complexity of the class considering the number of methods and amount of communication with other classes. It is a direct indicator of design complexity and testability |
| DIT | 3, Maximum 6 | It is the depth of inheritance hierarchy. High value is good for Reusability, but it will increase complexity. |

## II. EFFECT OF PROPERTIES OF OO PARADIGM ON DEFECT-PRONENESS OF THE SYSTEM

Our research is mainly focusing on analysis of properties [1] of OO paradigm that influences the desirable external quality factors and determines the entire software quality. In this study the contribution of internal design quality factors inheritance, class complexity and message passing or communication among objects on defect proneness is measured based on heuristics, the domain knowledge of experts in the field, our intuition, and with characteristics of CK Metrics. Table II depicts the empirical data on effect of design characteristics on defect-proneness of the system. Figure 1 is the graphical representation of data in Table II.

*Hypothesis ($H1_d$): Classes with high DIT values are associated with greater class complexity and fault-proneness.*

It has been reported widely that 95 to 92 % defect-proneness of the class in theoretical mode when DIT becomes 6 and this level is recommended as the maximum level for DIT. The properties originated during the interval of DIT between 2 and 4 are about 18 and 64 %. In the range of 5 to 6 the value is around 81% to 92%. For 0 depth of inheritance, with root class alone the fault-proneness will be negligible, because of less class complexity. The influence factors are shown in Table II and in the corresponding graph in figure 1.

TABLE II. EFFECT OF DESIGN CHARACTERISTICS ON DEFECT-DENSITY

| Design Characteristics-Metrics | | | | | |
|---|---|---|---|---|---|
| DIT ɭ | % Influence | RFC ȕ | % Influence | WMC Û | % Influence |
| 1 | 9 | 1 | 3 | 1 | 5 |
| 2 | 21 | 23 | 31 | 5 | 25 |
| 3 | 36 | 45 | 56 | 10 | 42 |
| 4 | 56 | 60 | 80 | 12 | 63 |
| 5 | 74 | 80 | 91 | 16 | 80 |
| 6 | 98 | 100 | 78 | 20 | 92 |
| | | 120 | 62 | 30 | 75 |
| | | 149 | 51 | 50 | 54 |
| | | 170 | 40 | 60 | 40 |
| | | 198 | 26 | 70 | 10 |
| | | 222 | 07 | 96 | 02 |

*Hypothesis 2($H2_d$): Classes with higher value of RFC are associated with high design complexity and fault-proneness.*

From industrial test data, it is observed that at desirable threshold of 50 the influence on defect-proneness of the class will be around 22%. When RFC increases merging of classes will take place, this leads to increased complexity of the system. Beyond the desired value at 100 RFC the defect-proneness of the class is reported to about 57%.With empirical analysis and the domain knowledge, it is observed that the fault-proneness will be increasing to a maximum of 92 to 100% for a value of 200 to 222 of RFC as shown in figure 1.

*Hypothesis3($H3_d$): Classes with higher values of WMC will be associated with high level of class complexity and more vulnerable to defects and classes with lower values will be associated with less defect density.*

From the experimental data, it is observed that the defect free classes are possible when WMC is at much less than the optimum level. It is observed that 2 to 9% influence on defect proneness when the class is having 1 to 5 methods in it. It is evident from figure 1 which is drawn based on the experimental data and the domain knowledge of experts, the defect proneness is increased when WMC metric value is increased. It is in the range of 13% to 18% for a value of 10 to 40 of the metric. There will be a high rise in defect proneness from 18% to 40% to 95% when WMC crosses the optimum value, as shown in Table II. This indicates the design is of highly complicated and containing more application specific modules which are more error prone and requires splitting of classes or modular design.

## III. MODEL EVALUATION AND EMPIRICAL VALIDATION

In our research, the empirical measurements and observations are conducted on 20 small and medium level real time commercial projects in various software industries including in-house industry (incubated software development industry in our research centre). In this paper we are presenting the estimated defect-proneness index through our model for five sample projects developed in Java platform. The computed results with our estimation model are shown in table III. Table II depicts the influence factors of design metrics arrived through well understanding of property influence on defect density. Suitable functional equations are developed for selected CK Metrics through multi functional estimation technique as shown in equations 1, 2 and 3. The Weighted Linear Combination of this functional estimation model is formulated through property based analysis (feature identification process), domain knowledge of experts (software reuse guidelines) and the empirical results from large scale industries on various real time projects and is shown in equation 4. Defect-proneness can be estimated through this model when design complexity metrics for each class component is given as input data. Figure 2 depicts the functional estimation model for the three structural metrics such as DIT (DPɭ), RFC ($DP_{ȕ}$) and WMC (DP Û).

$$DPɭ = -1.55 * ɭ^3 + 16.86 * ɭ^2 - 35.3 * ɭ + 30.67 \quad --- (1)$$

$$DP_{ȕ} = 0.002 * ȕ^2 + 0.06 * ȕ + 0.33 \quad --- (2)$$

$$DPÛ = 4.3*10\wedge(-3) * Û^{2+} 0.4 * Û + 4.3 \quad --- (3)$$

Figure 2. Functional estimation model for metrics

Here DTT = ⌊, RFC = ǔ and WMC = Û   for equations 1, 2, 3 and 4.

$$DPT \quad 0.25\,(1/n)\sum_{k=1}^{n}(DF_{D})_{k} + 0.37(1/n)\sum_{k=1}^{n}(DF_{E})_{k} + 0.38\,(1/n)\sum_{k=1}^{n}(DF_{y})_{k} \tag{4}$$

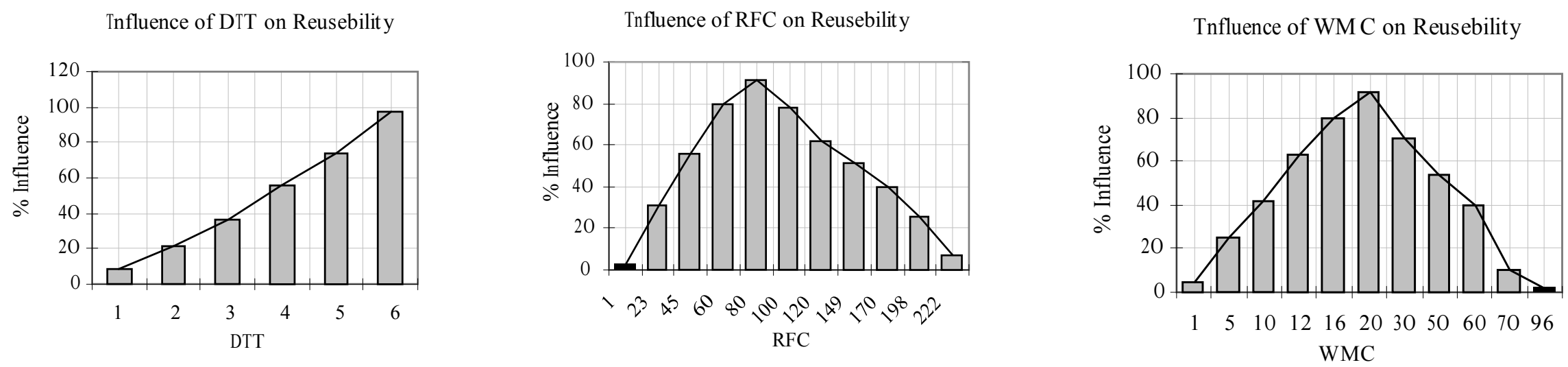

Figure 1. Tnfluence of Selected CK Metrics

Hitherto not much work is done to predict the defect-proneness index based on property analysis and we attempt to produce the functional interrelationship measurements and prediction. Tn the recent past many researchers made an attempt on the property analysis of OO characteristics and their impact on quality and certain features has been evolved. We have studied further and developed a novel estimation model which identifies the influence of design features on external quality characteristics of OOSS. This estimation model will provide an appropriate feedback on fault-proneness to designers and developers and help them to choose a better design and implementation plan that satisfies requirement of company and customer. This model is evolved through the weighted linear combination of the Functional Estimation equations; it is effectively used for direct computation of the defect-proneness index of any project represented by class level design measurements (CK Metrics Data input).

The impact of selected properties (DTT, RFC and WMC) of OO paradigm on defect-proneness is represented in the scale of 1. We have carried out a property based analysis on the influence of these properties on this quality factor to arrive at the weighted contribution factors of individual metrics. Our findings are validated with quality managers of two software industries, and also tested in our research laboratory.

The code level design complexity measurement of two commercial projects developed by software development industry incubated in our research centre is collected through an automated tool. Six developers having three years experience each, made involved in developing these projects using client server architecture, database technology for designing and Java code converter management. Table TTT shows the design metric measurement at code level for sample projects and Table TV shows the Defect-proneness index at class level and system level. Our result is verified with industry standards for defect level. Our estimation technique provides a result with 3 to 6 % of standard deviation from the test data.

We believe that our model will have significant contribution in software quality control activities to achieve better design for less defect prone classes and high-quality software products.

## TV. DTSCUSSTON OF RESULTS

The results of our model are depicted in Table TV for our sample projects. These results indicate the defect-proneness of the commercial projects developed on Java platform, at both class level and project level. The contribution of inheritance property for this quality factor for these samples is shown in the column DPDTT. The empirical data presented supports our hypothesis $H1_d$. Tt is observed that when the depth of inheritance increased to 4, the influence is around 57%. The influence of DTT on defect proneness is 1O% to 33% for a value of 1 to 3 from the root. From the experimental data it is evident that RFC will increase the defect proneness of the system for values above 1OO. For a higher value of 93 of RFC in our sample projects its contribution is around 24%. For a value of 24 of RFC, its influence is only 3%. When the communication or method invocation is in the reduced order in any projects, the complexity of the code will be less which will result in better quality project. The influence of RFC is shown as RURFC in the table. Tts influence will be around 48% to 9O% for a RFC value of 14O to 200. Thus it supports our hypothesis $H2_d$. When number of methods increases in a class, it will become more complex and does not support the quality of the system. The influence of WMC is varied between 17% and 44% on defect-proneness of the system for a value of 25 and 60 methods per class. For a lower value of 5 methods per class it is about 6%. When WMC increases beyond 30 the complexity of the system increases incredibly and the defect level also sky-scraping. Tt is observed from the sample 5, for 68 methods in class 2 it is observed that 51% of possible defects in the system is because of this poor design. Thus the empirical data presented is supporting our hypothesis $H3_d$. From the analysis of Java samples based on our model, it is observed that the contribution of RFC, WMC and DTT plays a vital role in determining the defect-proneness index of individual classes and projects.

Tt is observed from our experimental results that the contribution of these design properties of OO paradigm towards defect is minimum 8.23 % and maximum 44.18 % with a standard deviation of about 9.2%. These observations are verified with the actual defect data from industry and found that our estimation method gives a promising result and it can be further utilized for minimizing the defect proneness of the system with an appropriate adjustment of the design of the system.

## V. CONCLUSTON

Building a quality system has been the driving goal of all software engineering efforts over few decades. The lack of design and implementation guidance may affect the overall quality of the system which depends on reusability, defect level and maintainability of the system. A significant research effort is required to define quality measures. Measuring the structural design properties of software artifacts with design metrics, is a promising approach at an early stage. Our estimation model provides an assessment of the defect proneness of the system in an early stage by analyzing the interrelationship among the defect occurrence and design parameters of the software.

To our knowledge, the prior research on modeling of interaction between the design metrics and the defect proneness of classes and products are not much visible. The effects of design complexity metrics such as number of methods (WMC), inheritance depth (DTT),and response for message (RFC), on defect-proneness are found to have non linear influence as shown in table TT. Our future research include the analysis of other three metrics Viz. LCOM, CBO and NOC in the CK metric suite and the dynamic complexity measurements of object oriented paradigm on defect-proneness index at design level of the system.

TABLE TTT. DESTGN METRTCS TABLE TV. % DEFECT PRONENESS

| PROJECT | class | DTT | RFC | WMC | DPDTT | DPRFC | DPWMC | C-DPR |
|---|---|---|---|---|---|---|---|---|
| | 1 | 1 | 10 | 7 | 10.63 | 1.12 | 7.31 | 6.14 |
| | 2 | 3 | 9 | 3 | 33.31 | 1.02 | 5.54 | 11.03 |
| | 3 | 2 | 21 | 1 | 14.71 | 2.47 | 4.70 | 6.57 |
| | 4 | 2 | 3 | 9 | 14.71 | 0.52 | 8.25 | 7.34 |
| T | 5 | 1 | 1 | 7 | 10.63 | 0.39 | 7.31 | 5.87 |
| | | | | | | | | P-DPR |
| | | | | | 16.80 | 1.11 | 6.62 | 7.39 |
| | 1 | 5 | 24 | 14 | 75.67 | 2.93 | 10.74 | 24.51 |
| | 2 | 3 | 28 | 8 | 33.31 | 3.60 | 7.78 | 12.93 |
| TT | 3 | 1 | 27 | 13 | 10.63 | 3.43 | 10.23 | 8.22 |
| | 4 | 3 | 26 | 1 | 33.31 | 3.26 | 4.70 | 11.51 |
| | | | | | | | | P-DPR |
| | | | | | 38.23 | 3.30 | 8.36 | 14.29 |
| | 1 | 4 | 19 | 3 | 56.83 | 2.19 | 5.54 | 17.34 |
| | 2 | 2 | 27 | 5 | 14.71 | 3.43 | 6.41 | 7.64 |
| | 3 | 3 | 15 | 12 | 33.31 | 1.67 | 9.72 | 13.03 |
| | 4 | 4 | 27 | 9 | 56.83 | 3.43 | 8.25 | 18.94 |
| TTT | 5 | 3 | 23 | 4 | 33.31 | 2.77 | 5.97 | 11.86 |
| | | | | | | | | P-DPR |
| | | | | | 39.00 | 2.70 | 7.18 | 13.76 |
| | 1 | 6 | 31 | 6 | 80.23 | 4.15 | 16.99 | 28.73 |
| | 2 | 4 | 7 | 10 | 56.83 | 0.84 | 51.38 | 36.10 |
| | 3 | 4 | 27 | 1 | 56.83 | 3.43 | 40.18 | 32.35 |
| | 4 | 2 | 25 | 9 | 14.71 | 3.09 | 22.87 | 14.43 |
| TV | 5 | 3 | 6 | 8 | 33.31 | 0.75 | 6.41 | 11.30 |
| | 6 | 1 | 34 | 3 | 10.63 | 4.73 | 12.89 | 9.82 |
| | 7 | 3 | 12 | 2 | 33.31 | 1.33 | 43.78 | 27.21 |
| | 8 | 2 | 4 | 15 | 14.71 | 0.60 | 22.87 | 13.50 |
| | 9 | 5 | 35 | 6 | 75.67 | 4.93 | 15.77 | 27.37 |
| | 10 | 4 | 26 | 14 | 56.83 | 3.26 | 6.85 | 18.29 |
| | 11 | 5 | 11 | 8 | 75.67 | 1.22 | 7.78 | 22.64 |
| | | | | | | | | P-DPR |
| | | | | | 46.25 | 2.57 | 22.53 | 21.98 |
| | 1 | 1 | 14 | 25 | 10.63 | 1.55 | 16.99 | 10.37 |
| | 2 | 2 | 45 | 68 | 14.71 | 7.19 | 51.38 | 27.92 |
| | 3 | 2 | 52 | 56 | 14.71 | 9.02 | 40.18 | 23.89 |
| | 4 | 2 | 29 | 34 | 14.71 | 3.78 | 22.87 | 14.68 |
| V | 5 | 4 | 12 | 5 | 56.83 | 1.33 | 6.41 | 17.39 |
| | 6 | 4 | 11 | 18 | 56.83 | 1.22 | 12.89 | 20.07 |
| | 7 | 5 | 78 | 60 | 75.67 | 17.63 | 43.78 | 43.83 |
| | 8 | 1 | 93 | 34 | 10.63 | 23.89 | 22.87 | 21.10 |
| | 9 | 3 | 55 | 23 | 33.31 | 9.87 | 15.77 | 18.61 |
| | 10 | 2 | 46 | 6 | 14.71 | 7.44 | 6.85 | 9.31 |
| | | | | | | | | P-DPR |
| | | | | | 30.27 | 8.29 | 24.00 | 20.72 |

DPDTT( $DF_D$ ) – Defect Proneness due to DTT

DPRFC ( $DF_E$ ) – Defect Proneness due to RFC

DPWMC ( $DFy$ ) – Defect Proneness due to WMC

C-DPR – Defect Proneness at class level

P-DPR – Defect Proneness at Project level